\RequirePackage{lineno}

\documentclass[prc,reprint,twocolumn,showpacs,preprintnumbers,superscriptaddress,amsmath,amssymb,floatfix]{revtex4-1}

\pdfoutput=1

\usepackage{float}
\usepackage{graphicx}
\usepackage{dcolumn}
\usepackage{multirow}
\usepackage{bm}
\usepackage{epstopdf}
\usepackage{color}

\usepackage{helvet}

\usepackage{afterpage}

\newcommand{\mytext}[1]{\text{ #1}}

\newcommand{\GeV}{\ensuremath{\mytext{GeV}}}
\newcommand{\GeVc}{\ensuremath{\mytext{GeV}/c}}

\newcommand{\dif}{\text{d}}

\begin{document}

\title{Di-Hadron Correlations with Identified Leading Hadrons in 200 GeV Au+Au and d+Au Collisions at STAR}

\affiliation{AGH University of Science and Technology, Cracow 30-059, Poland}
\affiliation{Argonne National Laboratory, Argonne, Illinois 60439, USA}
\affiliation{Brookhaven National Laboratory, Upton, New York 11973, USA}
\affiliation{University of California, Berkeley, California 94720, USA}
\affiliation{University of California, Davis, California 95616, USA}
\affiliation{University of California, Los Angeles, California 90095, USA}
\affiliation{Central China Normal University (HZNU), Wuhan 430079, China}
\affiliation{University of Illinois at Chicago, Chicago, Illinois 60607, USA}
\affiliation{Creighton University, Omaha, Nebraska 68178, USA}
\affiliation{Czech Technical University in Prague, FNSPE, Prague, 115 19, Czech Republic}
\affiliation{Nuclear Physics Institute AS CR, 250 68 \v{R}e\v{z}/Prague, Czech Republic}
\affiliation{Frankfurt Institute for Advanced Studies FIAS, Frankfurt 60438, Germany}
\affiliation{Institute of Physics, Bhubaneswar 751005, India}
\affiliation{Indian Institute of Technology, Mumbai 400076, India}
\affiliation{Indiana University, Bloomington, Indiana 47408, USA}
\affiliation{Alikhanov Institute for Theoretical and Experimental Physics, Moscow 117218, Russia}
\affiliation{University of Jammu, Jammu 180001, India}
\affiliation{Joint Institute for Nuclear Research, Dubna, 141 980, Russia}
\affiliation{Kent State University, Kent, Ohio 44242, USA}
\affiliation{University of Kentucky, Lexington, Kentucky, 40506-0055, USA}
\affiliation{Korea Institute of Science and Technology Information, Daejeon 305-701, Korea}
\affiliation{Institute of Modern Physics, Lanzhou 730000, China}
\affiliation{Lawrence Berkeley National Laboratory, Berkeley, California 94720, USA}
\affiliation{Max-Planck-Institut fur Physik, Munich 80805, Germany}
\affiliation{Michigan State University, East Lansing, Michigan 48824, USA}
\affiliation{Moscow Engineering Physics Institute, Moscow 115409, Russia}
\affiliation{National Institute of Science Education and Research, Bhubaneswar 751005, India}
\affiliation{Ohio State University, Columbus, Ohio 43210, USA}
\affiliation{Institute of Nuclear Physics PAN, Cracow 31-342, Poland}
\affiliation{Panjab University, Chandigarh 160014, India}
\affiliation{Pennsylvania State University, University Park, Pennsylvania 16802, USA}
\affiliation{Institute of High Energy Physics, Protvino 142281, Russia}
\affiliation{Purdue University, West Lafayette, Indiana 47907, USA}
\affiliation{Pusan National University, Pusan 609735, Republic of Korea}
\affiliation{University of Rajasthan, Jaipur 302004, India}
\affiliation{Rice University, Houston, Texas 77251, USA}
\affiliation{University of Science and Technology of China, Hefei 230026, China}
\affiliation{Shandong University, Jinan, Shandong 250100, China}
\affiliation{Shanghai Institute of Applied Physics, Shanghai 201800, China}
\affiliation{Temple University, Philadelphia, Pennsylvania 19122, USA}
\affiliation{Texas A\&M University, College Station, Texas 77843, USA}
\affiliation{University of Texas, Austin, Texas 78712, USA}
\affiliation{University of Houston, Houston, Texas 77204, USA}
\affiliation{Tsinghua University, Beijing 100084, China}
\affiliation{United States Naval Academy, Annapolis, Maryland, 21402, USA}
\affiliation{Valparaiso University, Valparaiso, Indiana 46383, USA}
\affiliation{Variable Energy Cyclotron Centre, Kolkata 700064, India}
\affiliation{Warsaw University of Technology, Warsaw 00-661, Poland}
\affiliation{Wayne State University, Detroit, Michigan 48201, USA}
\affiliation{World Laboratory for Cosmology and Particle Physics (WLCAPP), Cairo 11571, Egypt}
\affiliation{Yale University, New Haven, Connecticut 06520, USA}
\affiliation{University of Zagreb, Zagreb, HR-10002, Croatia}

\author{L.~Adamczyk}\affiliation{AGH University of Science and Technology, Cracow 30-059, Poland}
\author{J.~K.~Adkins}\affiliation{University of Kentucky, Lexington, Kentucky, 40506-0055, USA}
\author{G.~Agakishiev}\affiliation{Joint Institute for Nuclear Research, Dubna, 141 980, Russia}
\author{M.~M.~Aggarwal}\affiliation{Panjab University, Chandigarh 160014, India}
\author{Z.~Ahammed}\affiliation{Variable Energy Cyclotron Centre, Kolkata 700064, India}
\author{I.~Alekseev}\affiliation{Alikhanov Institute for Theoretical and Experimental Physics, Moscow 117218, Russia}
\author{J.~Alford}\affiliation{Kent State University, Kent, Ohio 44242, USA}
\author{A.~Aparin}\affiliation{Joint Institute for Nuclear Research, Dubna, 141 980, Russia}
\author{D.~Arkhipkin}\affiliation{Brookhaven National Laboratory, Upton, New York 11973, USA}
\author{E.~C.~Aschenauer}\affiliation{Brookhaven National Laboratory, Upton, New York 11973, USA}
% ...

\collaboration{STAR Collaboration}\noaffiliation


\date{September 30, 2014  -- Version 18}

\begin{abstract}
The STAR collaboration presents for the first time two-dimensional di-hadron correlations with identified leading hadrons in
200 GeV central Au+Au and minimum-bias d+Au collisions to explore hadronization mechanisms in the quark gluon plasma.
The enhancement of the jet-like yield for leading pions in Au+Au data with respect to the d+Au reference 
and the absence of such an enhancement for leading non-pions (protons and kaons) are discussed 
within the context of a quark recombination scenario.
The correlated yield at large angles, specifically in the \emph{ridge region}, is found to be significantly higher 
for leading non-pions than pions.
The consistencies of the constituent quark scaling, azimuthal harmonic model and a mini-jet modification model description of
the data are tested, providing further constraints on hadronization.
\end{abstract}


\maketitle

Experimental data from  heavy-ion collisions at ultra-relativistic energies achieved at the Relativistic Heavy Ion Collider (RHIC), and more recently at the Large Hadron Collider (LHC), are
conventionally interpreted in terms of a  unique form of matter,  the strongly-interacting Quark Gluon Plasma (sQGP).
It is estimated that temperatures reached in those collisions~\cite{PHENIX_temp,ALICE_temp} are well above the critical values predicted by lattice quantum chromodynamics calculations for the phase transition between hadronic and de-confined (partonic) matter~\cite{Karsch2002199}.
The RHIC experiments concluded that the formed medium displays the
properties of a nearly perfect liquid~\cite{Adams:2005dq,*Adcox:2004mh,*Back:2004je,*Arsene:2004fa}. 
A distinct feature of the sQGP
is jet quenching, which describes the large energy loss of ``hard'' (high transverse momentum, $p_T$) probes observed for example
in measurements of inclusive hadron distributions~\cite{PhysRevLett.91.172302,*PhysRevLett.91.072303}.

Jet quenching is also evident in modifications of back-to-back di-hadron correlations with a leading
(high-$p_T$) ``trigger'' hadron in Au+Au collisions in comparison to 
p+p and d+Au data~\cite{Adams:2005ph,Abelev:Ridge,STAR:LongRidge,STAR:DihadronPaper,PhysRevLett.91.072304,*Adare:2006nr,*Adare:2008ae,*PhysRevC.83.061901,*PhysRevLett.105.022301}.
One of the striking features found in di-hadron correlations from heavy-ion collisions is the emergence of a long-range plateau in relative pseudorapidity ($\Delta\eta$) 
on the near-side of a trigger hadron (small relative azimuth $\Delta\phi$), referred to as the ``ridge''~\cite{Adams:2005ph,Abelev:Ridge,STAR:LongRidge}.
The majority of recent theoretical descriptions of this phenomenon
invoke a transport  of initial-state to final-state anisotropy
via hydrodynamic expansion,
thus connecting measured observables to transport coefficients
and properties of the medium~\cite{Alver:2010gr,*Alver:Erratum,Luzum:2010sp,Gale:2013da}. 
Most of the proposed alternative explanations also require hydrodynamic 
evolution of the medium to reproduce 
the ridge~\cite{Wong2011,*Hwa2011,*Andres2014,*Moschelli2010,*Petersen2011,*Zhang2013,*Dumitru2011}.
The latest observations of ridge-like correlations  in high-multiplicity p+p
and p+Pb collisions at the LHC provide new tests of theoretical 
explanations of the ridge~\cite{CMS:ppRidge,*CMS:pPbRidge,*Aad:2013fja,*ATLAS:ppRidge13,*Meissner:2015fha,*Adam:2015bka,Werner:2014xoa,*Dusling:2014oha,*Bautista:2015zqu,*Altinoluk:2015uaa}.
 
Another anomaly, the enhancement of the relative baryon-to-meson production, was discovered at RHIC in
the intermediate-$p_T$ range between 2 and 5 GeV/$c$,
where the ridge happens to be most prominent~\cite{STAR:BaryonMeson,PHENIX:BaryonMeson,STAR:LongBaryonMeson,Abelev:2014laa}.
The ratio of proton to pion yields in central Au+Au collisions exceeds by more than a factor of two 
that in d+Au and p+p events.
Similar baryon enhancements were reported in the strange-hadron sector~\cite{Aggarwal:2010ig,Abelev:2013xaa}.
In the same kinematic region, baryons and mesons exhibit different trends in azimuthal anisotropy, which at RHIC appear to
scale with the number of constituent quarks~\cite{STAR:NCQV2,*PHENIX:NCQV2}.
Recombination models, which incorporate the coalescence of two or three constituent quarks 
as a formation mechanism for mesons and baryons, are able to reproduce the observed enhancements 
in inclusive measurements~\cite{Hwa:Reco,*Fries:Reco}.
Description of hadronization processes remains challenging for theoretical calculations
(see for example  the unexpected measurements reported in Ref.~\cite{Airapetian:2007vu,*Airapetian:2011jp}); 
we expect these new measurements will facilitate further developments in this area.

In this Letter, we use angular correlations of intermediate-$p_T$ particles with identified leading hadrons
to further explore possible hadronization mechanisms in the quark gluon plasma, 
including changes to parton fragmentation patterns, 
dilution effects (reduction in per-trigger yields) due to recombination contributions, 
and quark number scaling behavior in correlations at large relative angles.

\begin{figure*}[!ht]
  \centering


  \setlength{\unitlength}{0.32\textwidth}
  \includegraphics[width=0.32\textwidth]{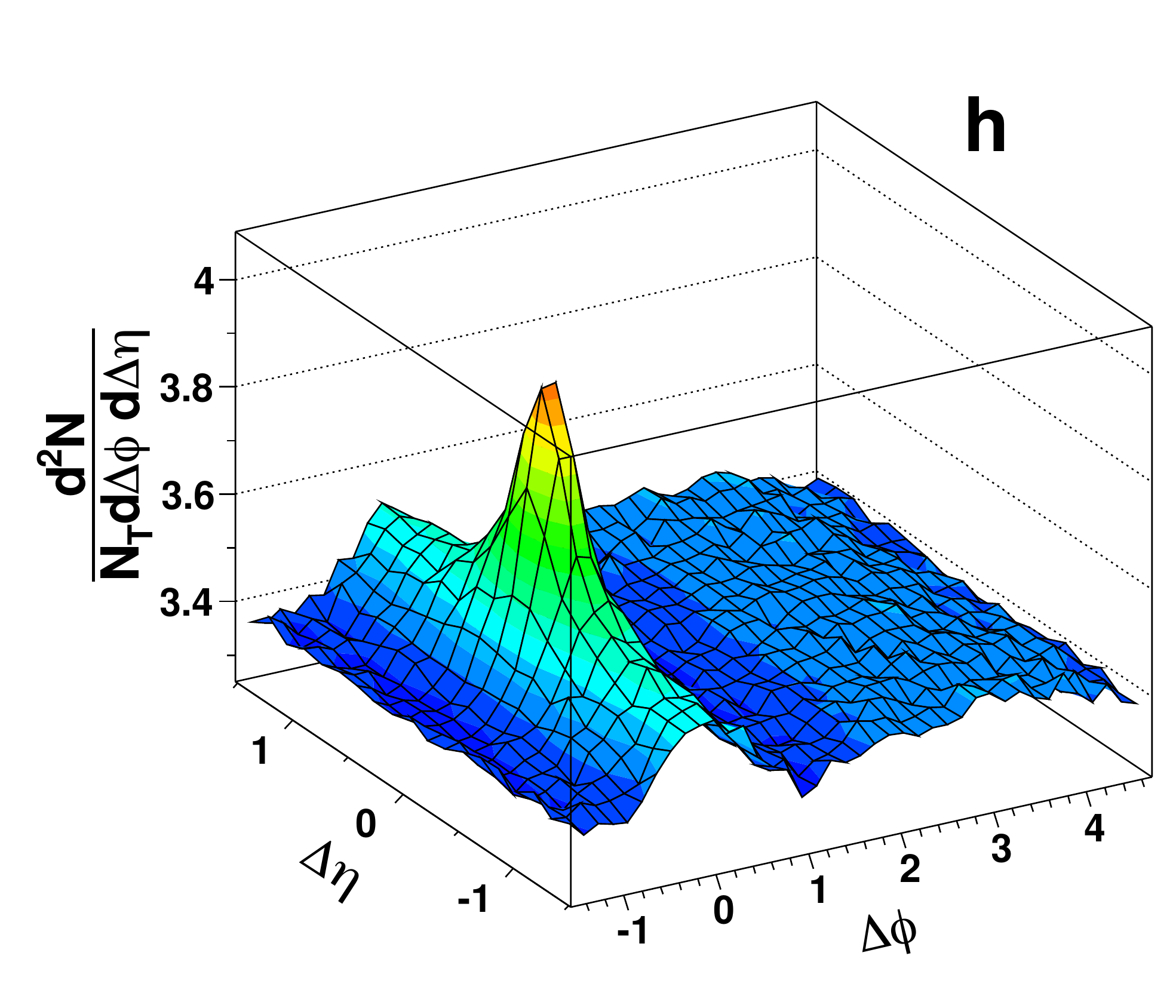}
  \put(-0.95,.75){\textbf{\textsf{Au+Au 0-10\%}}}
  \includegraphics[width=0.32\textwidth]{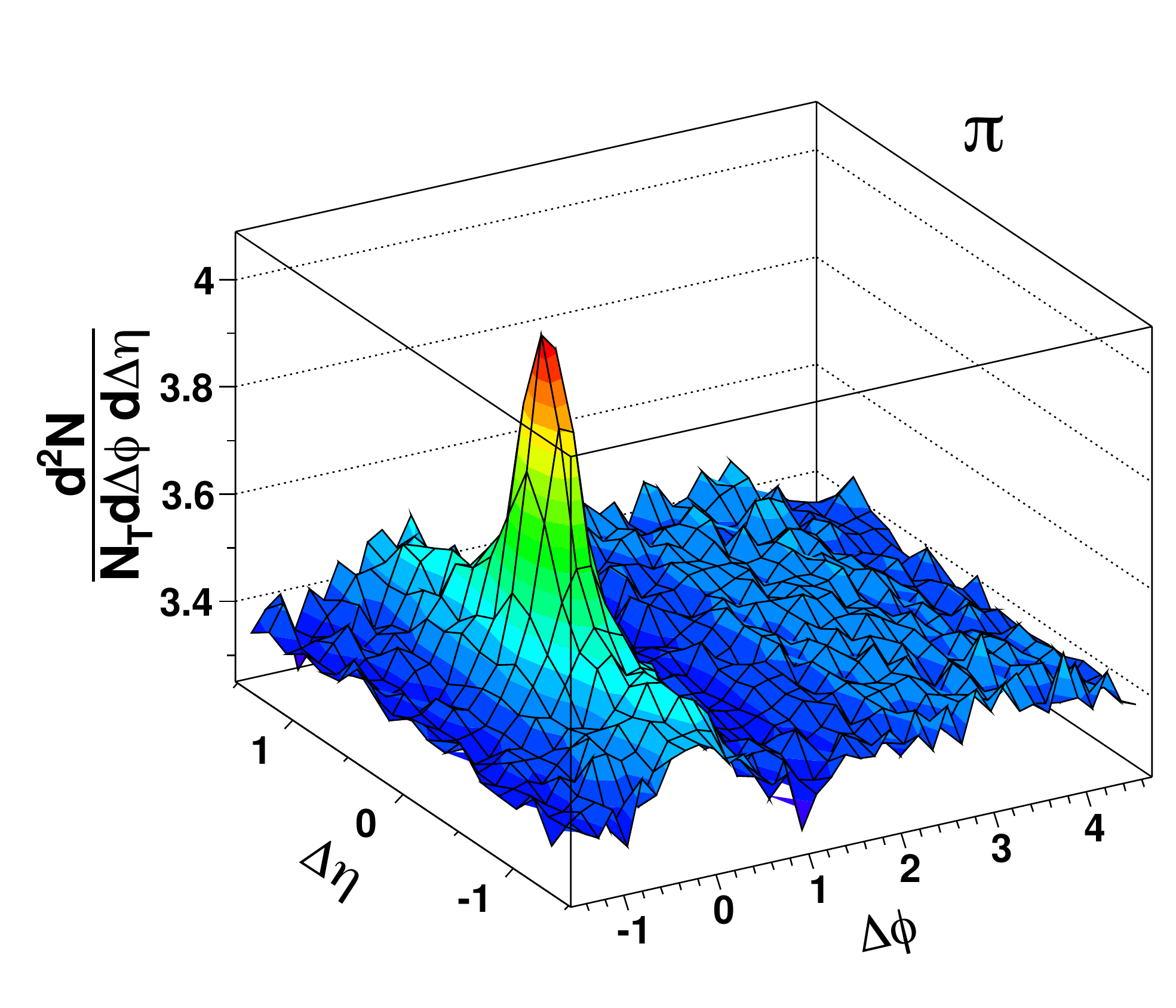}
  \includegraphics[width=0.32\textwidth]{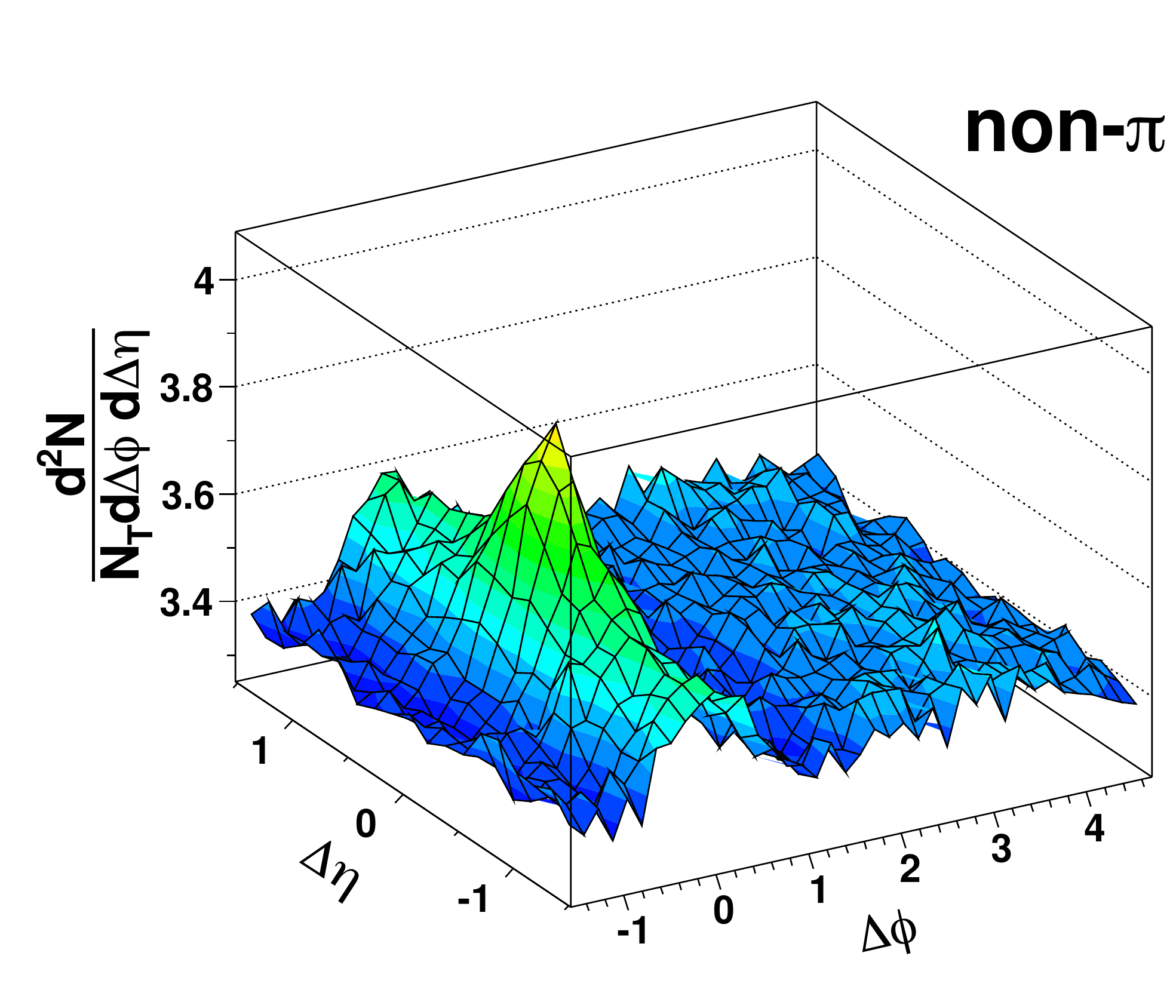}
  
  \includegraphics[width=0.32\textwidth]{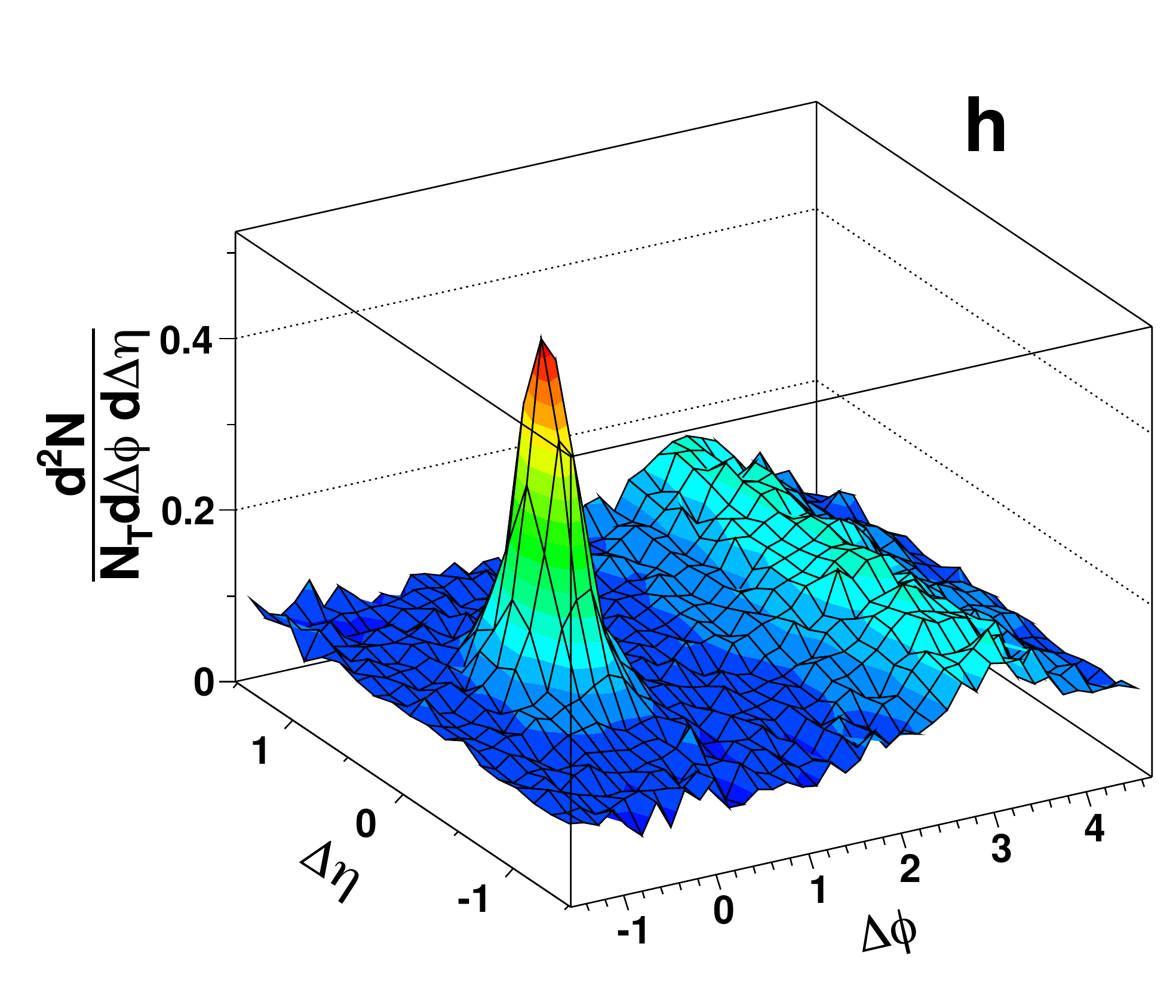}
  \put(-0.95,.75){\textbf{\textsf{d+Au MinBias}}}
  \includegraphics[width=0.32\textwidth]{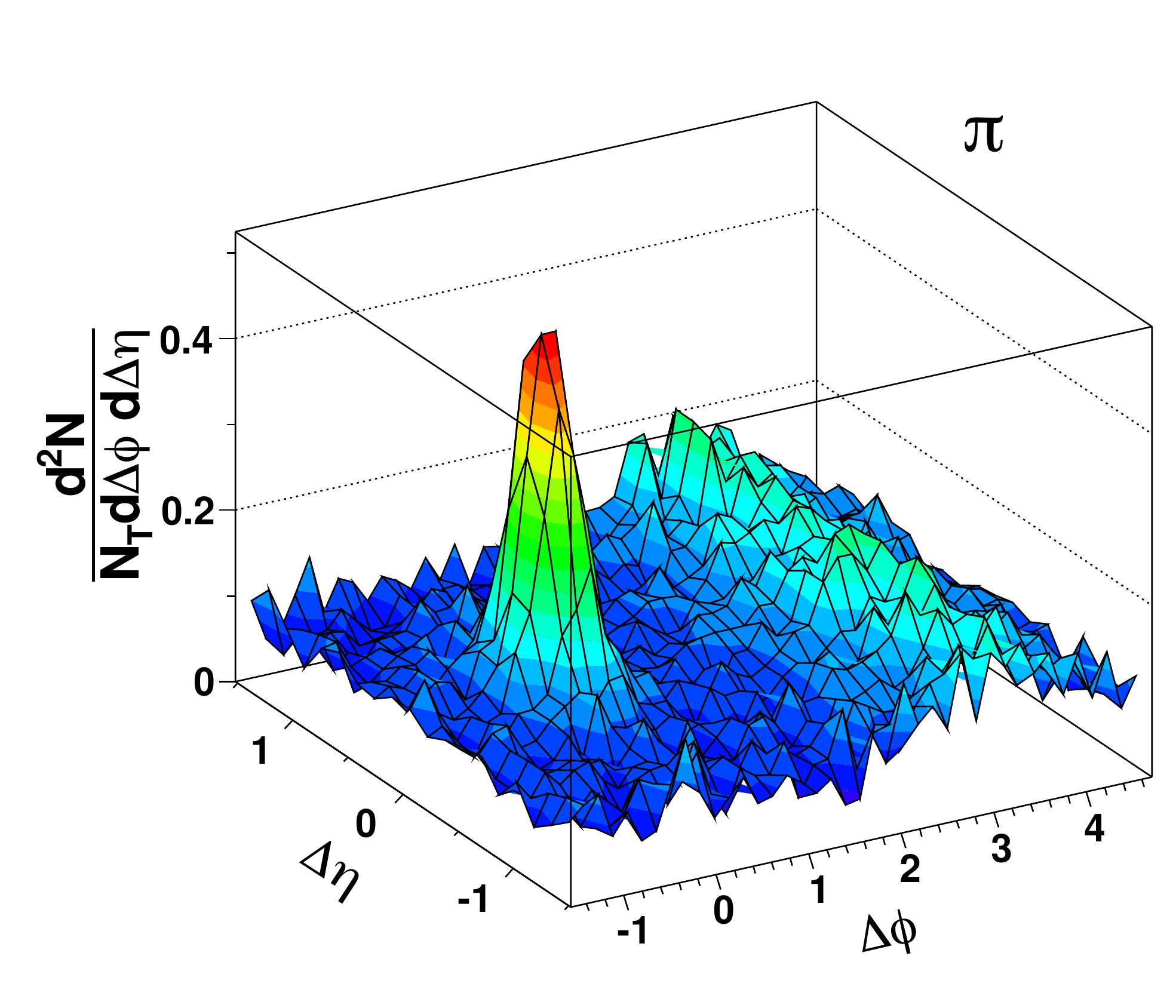}
  \includegraphics[width=0.32\textwidth]{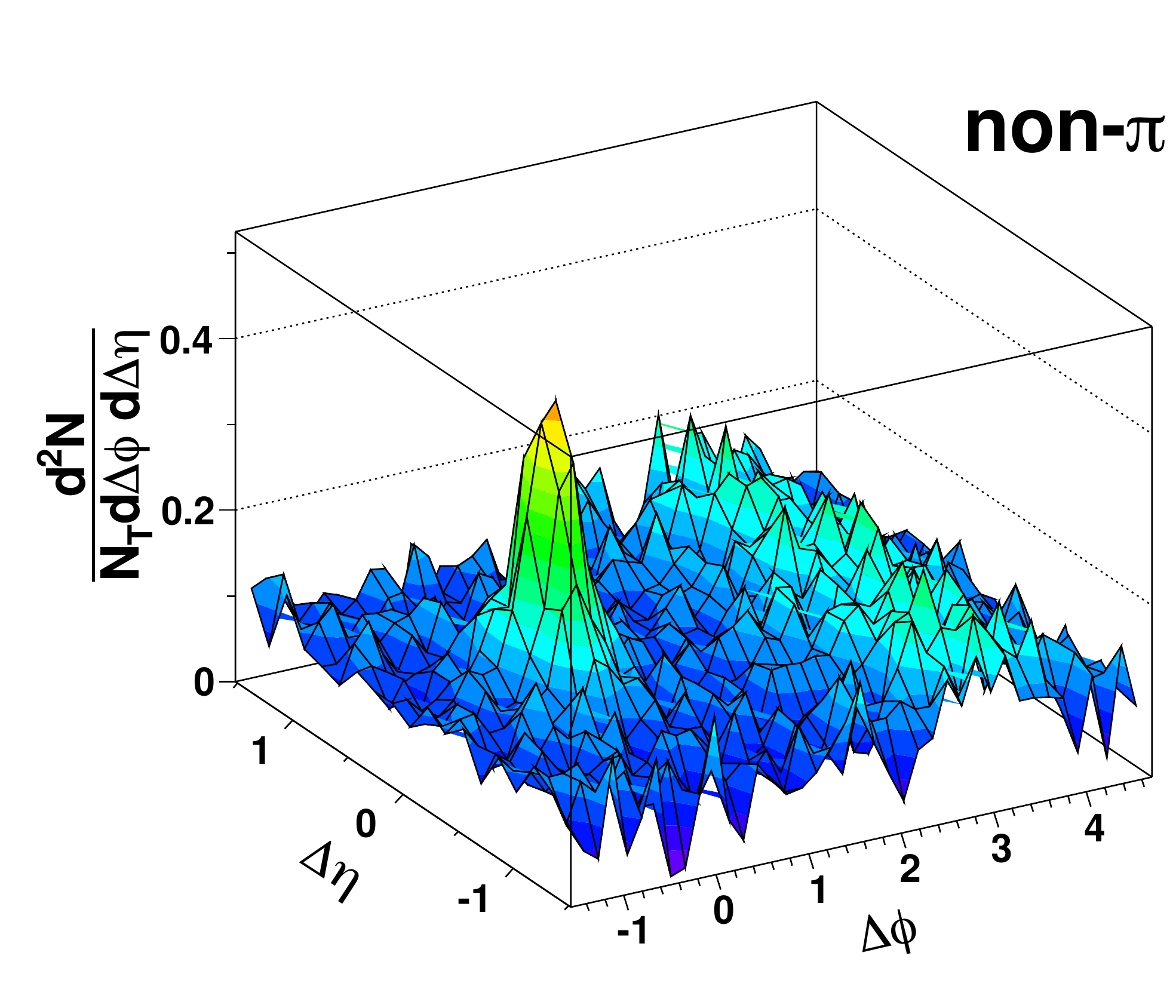}

  \caption{(Color online) Two-dimensional $\Delta\phi$ vs. $\Delta\eta$ correlation functions for charged hadron (left), pion (middle), 
    and non-pion (right) triggers from  0-10\% most-central Au+Au (top row) and minimum-bias d+Au (bottom row) data at 200~GeV.
    All trigger and associated charged hadrons  are selected in the respective $p_T$ ranges $4<p_T^{\text{trig}}<5~\GeVc$
    and $1.5<p_T^{\text{assoc}}<4$~\GeVc.
  }
  \label{fig:Raw2D}
\end{figure*}
Two-dimensional di-hadron correlations in $\Delta\phi$ and $\Delta\eta$, 
with statistically separated pion and non-pion triggers,
are studied for the 0-10\% most-central Au+Au and minimum-bias d+Au collisions at the center-of-mass energy per nucleon pair 
$\sqrt{s_{NN}}=200$ GeV.
No charge separation is considered; in this paper, the terms pion, proton, and kaon 
will be used to refer to the sum of particles and their respective anti-particles.
We separately study the short-range (jet-like peak) and long-range (ridge) correlations
in central Au+Au data, comparing to
reference measurements performed in d+Au collisions.
Details of the short-range correlations can shed light on the interplay between parton fragmentation, energy loss,
and recombination processes in the quark gluon plasma.
The long-range correlations are studied with two approaches:
(1) Fourier decomposition where we extract the azimuthal harmonic amplitudes which in some approaches are
interpreted as hydrodynamic ``flows''~\cite{Luzum:2010sp},
and (2) a mini-jet (defined in~\cite{STAR:Anomalous}) modification model~\cite{STAR:Anomalous,Ray2013}.

The analysis was conducted using $1.52\times10^8$ central-triggered Au+Au events at $\sqrt{s_{NN}}=200\GeV$  from STAR's 2010 data run,
and $4.6\times10^7$ events from the 2008 minimum-bias 200~GeV d+Au data set.
Particle densities as well as Glauber Monte Carlo results for these centrality selections can be found in Ref.~\cite{Abelev:2008ab}.
The STAR Time Projection Chamber (TPC)~\cite{Anderson:2003ur} was used for tracking, momentum reconstruction  and particle identification. 
Contamination by tracks from another collision (``pileup''), which  can distort the shape of di-hadron correlations~\cite{STAR:Anomalous},
was removed by rejecting events with an abnormally large (over three 
standard deviations above the average) number of tracks not originating from the primary vertex.

Trigger particles are defined as the highest-$p_T$ charged hadron in a given event
with $p_T^{\text{trig}}$ between 4 and 5~GeV/$c$; charged hadrons between 1.5 and 4 GeV/c are associated with each trigger particle.
This kinematic range focuses on a particularly interesting region where trigger particle production is thought to be dominated by fragmentation,
at least in $pp$ and d+Au collisions.
For the same range in Au+Au events, the baryon-to-meson enhancement is large, suggesting significant recombination contributions~\cite{Fries:NewReco,*Hwa:NewReco}.
Since the medium induced jet quenching affects the correlations in essentially an opposite way from thermal parton recombination contributions,
comparing the correlations for proton and pion triggers provides an additional handle for separating these effects.
Also, the ridge and away-side modifications in this $p_T$ range are significant, 
and elliptic flow, and its $p_T$ dependence, are minimal, thus facilitating the present tests of constituent quark number
scaling. 
All trigger particles are required to have at least 30 TPC points per track (for optimal identification),
otherwise, standard quality cuts and corrections are applied as described in ref.~\cite{STAR:LongRidge}.
After quality cuts, $3.5\times10^6$ Au+Au and $1\times10^5$ d+Au events with a trigger particle were used for the analysis.
The statistical hadron identification procedure relies on the measured ionization energy loss ($\dif{E}/\dif{x}$) in the TPC gas. 
The $\dif{E}/\dif{x}$ calibration was carried out individually for five pseudorapidity and two trigger $p_T$ bins.
Details of the particle identification (PID) technique are identical to those in refs.~\cite{STAR:BaryonMeson,Abelev:2008ab,Shao:2005iu}.

We construct a two-dimensional correlation with each trigger and all associated hadrons in an event,
following the procedure outlined in ref.~\cite{STAR:LongRidge}.
Pion identification is straightforward: selecting triggers with $\dif E / \dif x$ above
the central (expected) pion value provides a sample with 98\% pion purity and, by construction, 50\% selection efficiency.
The ``pure-pion'' correlation is constructed with those triggers.
The remaining triggers are comprised of all protons, about 97\% of all kaons, and the remaining 50\% of pions.

We remove the pion contribution from the correlation with those remaining triggers by
direct subtraction of the pure-pion-triggered correlation.
The resulting ``non-pion''  correlation is then associated with a mixture of proton and kaon triggers
(about three protons for every two kaons~\cite{Adams:2006wk,*Adams:2006nd}).
Separating kaons from protons is complicated by the small $\dif E / \dif x$ difference
between the two and was not attempted in this Letter.
The systematic uncertainty due to the pion subtraction procedure is included in the PID uncertainty.
The evaluation procedure is similar to previous identified particle analyses with the STAR Time Projection Chamber,
where sensitivities of the final observables to systematic variations in the $\dif E / \dif x$ cut parameters were determined.
The feed-down contribution from weak decay daughters to the trigger particles cannot be disentangled directly.
Due to decay kinematics, the dominant feed-down contribution originates from $\Lambda\to p\pi$
and is estimated to constitute about 5\% of the non-pion triggers.
Resonance contributions are greatly suppressed at high $p_T$ and shown to give only minor contributions to correlation structures~\cite{STAR:Anomalous}.
All raw correlation functions are  corrected for detector inefficiency derived from Monte Carlo tracks 
embedded into real data as in refs.~\cite{STAR:LongRidge,Abelev:2008ab,STAR:DihadronPaper}.
Pair-acceptance effects are corrected using the mixed-event technique as in ref.~\cite{STAR:DihadronPaper}. 
The resulting correlations are shown in Fig.~\ref{fig:Raw2D}, with visible differences 
between the two trigger types in both jet-like peak and large
$\Delta\eta$ region in Au+Au. A significantly larger ridge amplitude is seen  for non-pion triggers, 
while the jet-like peak is more pronounced for the pion triggers.
By comparison, the correlations in d+Au show no discernible ridge on
the near-side, while differences between trigger types in the jet-like region
are qualitatively similar to Au+Au, suggesting that these may be partly due to kinematic effects.
In the following, we analyze these modifications individually. 

\begin{figure*}[hbt]
  \centering
  \includegraphics[width=\textwidth]{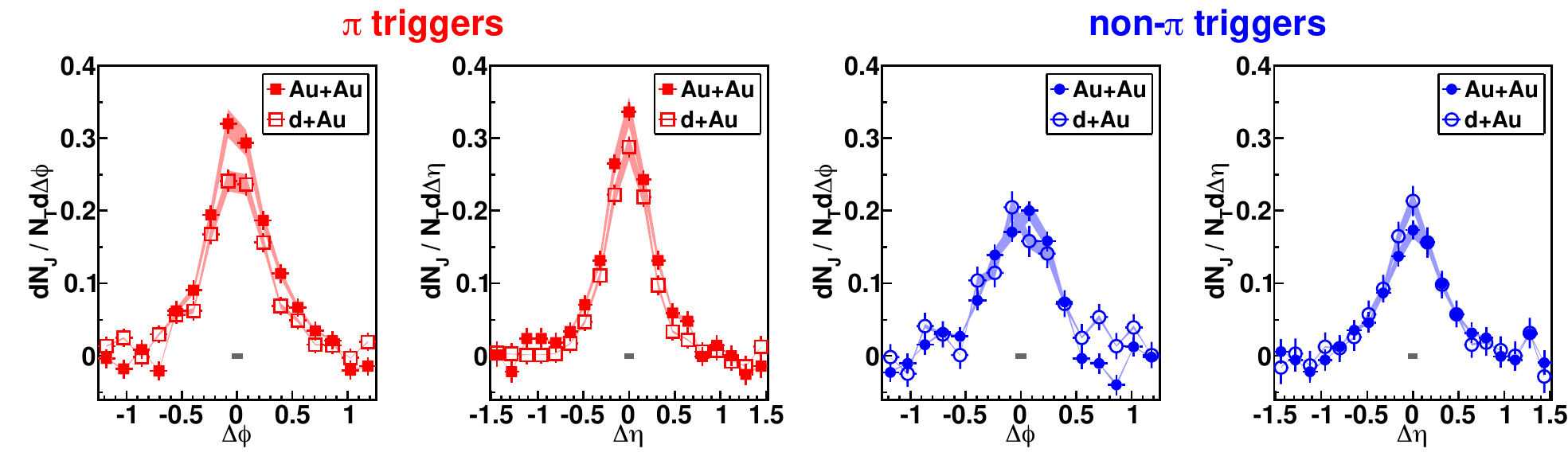}
  \caption{(Color online) 
    The $\Delta\phi$ and $\Delta\eta$ projections of the pure-cone correlations for $|\Delta\eta|<0.78$ and $|\Delta\phi|<\pi/4$, respectively,
    for pion triggers (left two panels) and non-pion triggers (right two panels). 
    Filled symbols show data  from the 0-10\% most-central Au+Au collisions at 200~GeV; 
    open symbols show data from minimum-bias d+Au data at the same energy. 
    Shaded boxes centered at zero show the uncertainty in background level determination; colored bands show the remaining systematic uncertainties.
  }
  \label{fig:Projection}
\end{figure*}

Initially, we study  the small-angle jet-like correlated signals. 
Assuming that all background contributions are $\Delta\eta$-independent,
as shown in refs.~\cite{STAR:LongRidge,Abelev2008}, 
we subtract those contributions averaged over large $|\Delta\eta|=$ 0.9--1.5 from the full correlations, resulting in ``pure-cone'' distributions. 
This procedure is supported by the two-dimensional fits to the data described below. 
We then calculate the fiducial jet-like yield in $|\Delta\eta|<0.78$, $|\Delta\phi|<\pi/4$ as in ref.~\cite{Abelev:Ridge}.
To isolate medium effects from initial-state nuclear effects, the Au+Au results are
then compared to the correlation function constructed  in an identical way for d+Au data (see Fig.~\ref{fig:Projection}). 
We report significant differences in the jet-like yield per trigger between the two systems for pion triggers.
At the same time,  correlations with non-pion triggers show, within uncertainties, similar yields for the two systems.
For quantitative comparisons, the integrated yields are presented in Table~\ref{tab:yields}. 
The yield extrapolation outside the fiducial range is performed using cone-shape modeling described below.
The systematic errors are dominated by the tracking efficiency uncertainty (5\%); other sources include  uncertainties from 
$p_T$ resolution ($3\%$), PID uncertainty ($2-3\%$), background level determination ($2\%$ for Au+Au, 2--5\% for d+Au;
found by varying the range for the $\Delta\eta$-independent ridge structure between $|\Delta\eta|=0.8\text{--}1.4$
and $|\Delta\eta|=1.0\text{--}1.6$ ),
track splitting/merging correction ($1\%$), and pair acceptance ($<$1\%).
The effect of feed-down protons on the jet-like yield is estimated to be less than 1\%.
\begin{table} [!htb]
  \centering
  \caption{Fiducial ($\left[|\Delta\eta|<0.78\right] \times \left[|\Delta\phi|<\pi/4\right]$) and extrapolated pure-cone yields
    for pion, non-pion and charged hadron (unidentified) triggers (see text), and the associated yield ratios.\label{tab:yields}}
  \begin{ruledtabular}
    \begin{tabular}{  c || c || c | c | c || c || c | c | c }
      {\bf Trigger } & \multicolumn{4}{ c|| }{\bf Au+Au 0-10\% } &\multicolumn{4}{ c } {\bf d+Au MinBias } \\ 
      \hline
                            &  Fid.   & Ext.    & Stat.   & Sys.  & Fid.    & Ext.    & Stat.   & Sys. \\ \hline
      $\pi$             & 0.211 & 0.214 & 3\% & 7\%  & 0.171 & 0.171 & 4\% & 6\% \\ \hline
      non-$\pi$     & 0.136 & 0.142 & 5\% & 6\%  & 0.142 & 0.148 & 7\% & 8\% \\ \hline
      All                 & 0.176  & 0.180 & 2\% & 5\%  & 0.161 & 0.168 & 2\% & 5\% \\ \hline
      \hline
      $\frac{\text{Y(non-}\pi)} {\text{Y(}\pi)}$ 
                           & 0.643  & 0.662 & 6\%  &  5\% & 0.835 & 0.866 & 8\% &  8\% \\ 
    \end{tabular}
  \end{ruledtabular}
\end{table}

The jet-like yield in the $p_T$ range 1.5-4~GeV/$c$ associated with  pion triggers in central Au+Au collisions is enhanced by 
$24\pm6 (\text{stat.}) \pm11 (\text{sys.})\%$ with respect to the reference measurement in d+Au. The yields for non-pion triggers are found to be similar between the two systems.
A previous work found similar trends in near-side associated yields~\cite{PHENIX:PidCorr};
however, in that one-dimensional analysis, no separation between jet-like peak and ridge contributions was possible.
We find that the jet-like yield for unidentified charged hadron triggers is also enhanced, consistent with our identified trigger results.
The enhancement of the jet-like yield of soft hadrons associated with pion triggers could be caused by the jet-quenching effect and/or 
medium-induced modification of fragmentation functions, and is qualitatively consistent with other observations 
from non-identified correlations~\cite{Aamodt2012} and direct jet 
measurements~\cite{Adamczyk:2013jei,Chatrchyan:2014ava,Aad:2014wha} for low $p_T$ hadrons. 
It is expected that a larger fraction of non-pion triggers are produced from gluon-jets
rather than quark-jets compared to pion triggers~\cite{Albino:AKK,Adams:2006nd,Mohanty2007}.
A predicted higher energy loss for in-medium gluons should then result in even larger jet-like yields 
for non-pion triggers~\cite{PhysRevC.58.2321,*PhysRevC.71.014903,*Wicks:2005gt,*Armesto:2005iq}.
On the other hand, particle production from recombination should produce smaller yields
than particle production from hard processes (fragmentation)~\cite{Fries:NewReco,*Hwa:NewReco},
thus diluting (reducing) per-trigger associated yields. This dilution effect would be stronger for baryons, as more intermediate-$p_T$ 
baryons than mesons are expected to be formed through such a mechanism.
The associated yields for non-pion triggers combine both of these competing effects.
Thus the observed reduction could be due to a larger recombination effect relative to that from the
increased energy loss expected for non-pion leading particles.
%

The ratio of associated yields for non-pion and pion leading hadrons
is shown in Table~\ref{tab:yields} for Au+Au and d+Au systems.
In these ratios, dominant contributions to systematic uncertainties
from the tracking efficiency estimate cancel out.
The double-ratio constructed from these two results quantifies the relative decrease in associated
jet-like yields for non-pion triggers with respect to leading pion results in Au+Au compared to d+Au.
This double-ratio, $0.76 \pm 0.08 (\text{stat.}) \pm 0.07 (\text{sys.})$,
can measure the net effect of the competition between higher energy loss/higher associated yields
at lower $p_T$ for  gluon jets versus reduced yields due to recombination in central Au+Au collisions.
Currently, no quantitative predictions for either of these two mechanisms are available for direct comparison with data.

\begin{figure*}[!hbt]
  \centering
  \includegraphics[width=0.32\textwidth]{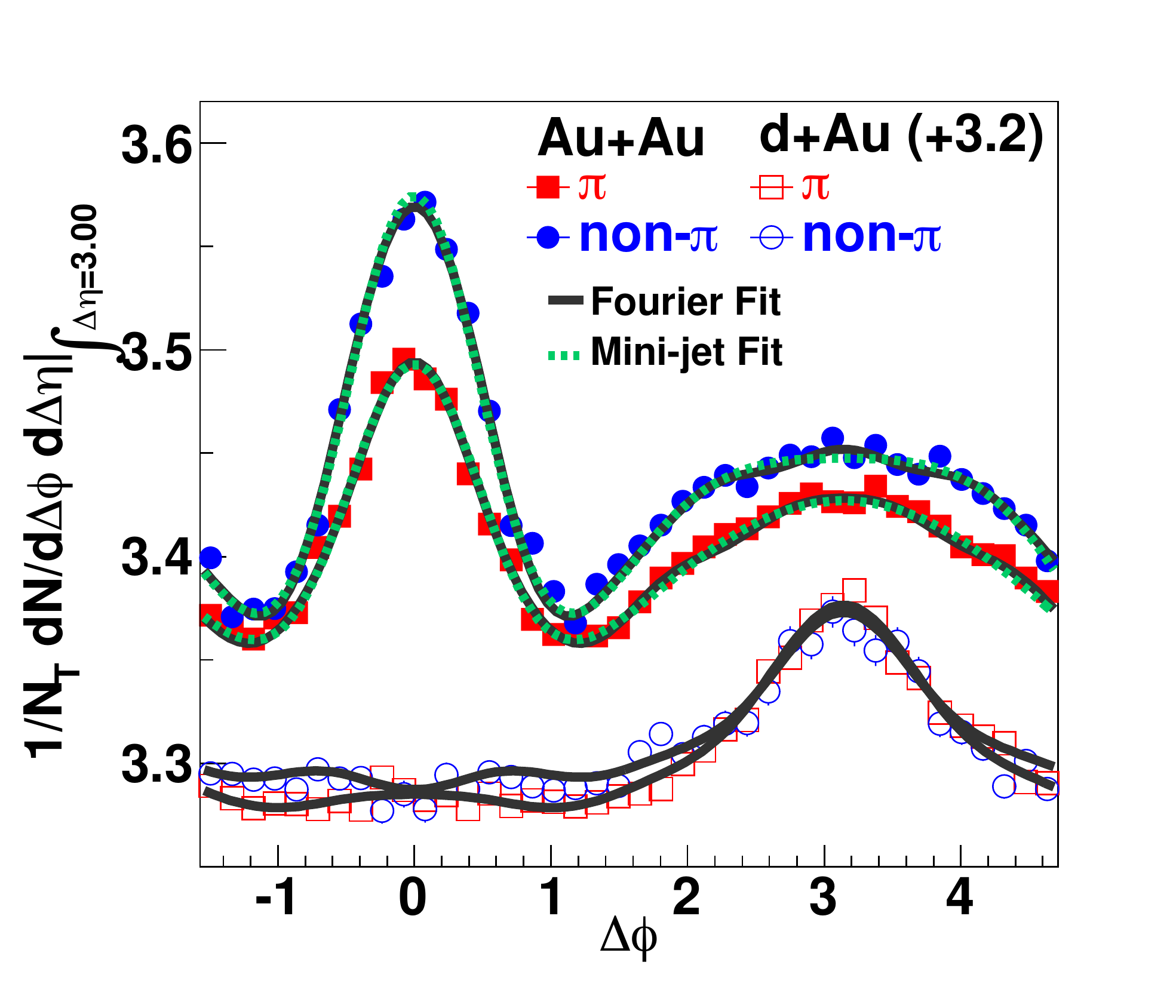}
  \includegraphics[width=0.32\textwidth]{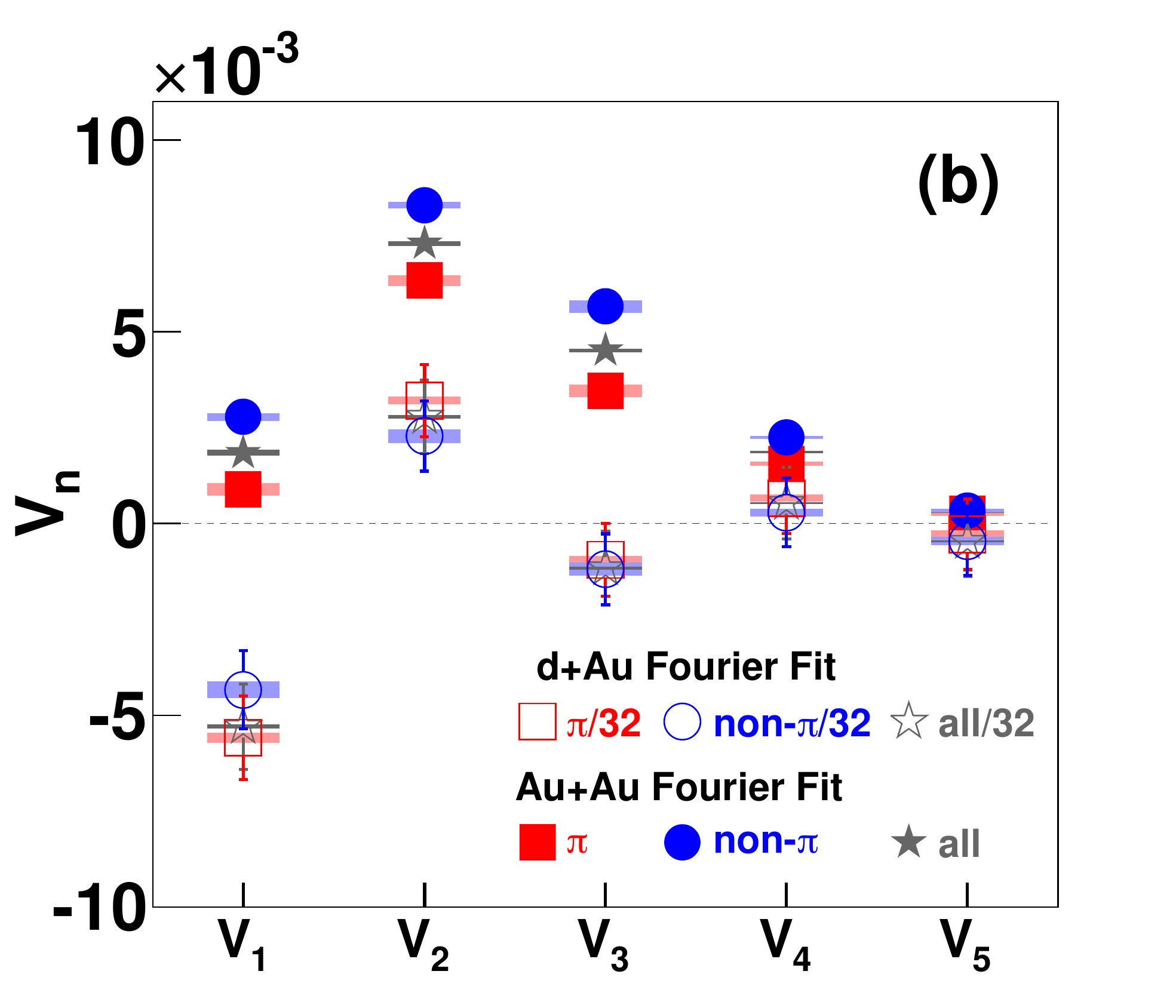}
  \includegraphics[width=0.32\textwidth]{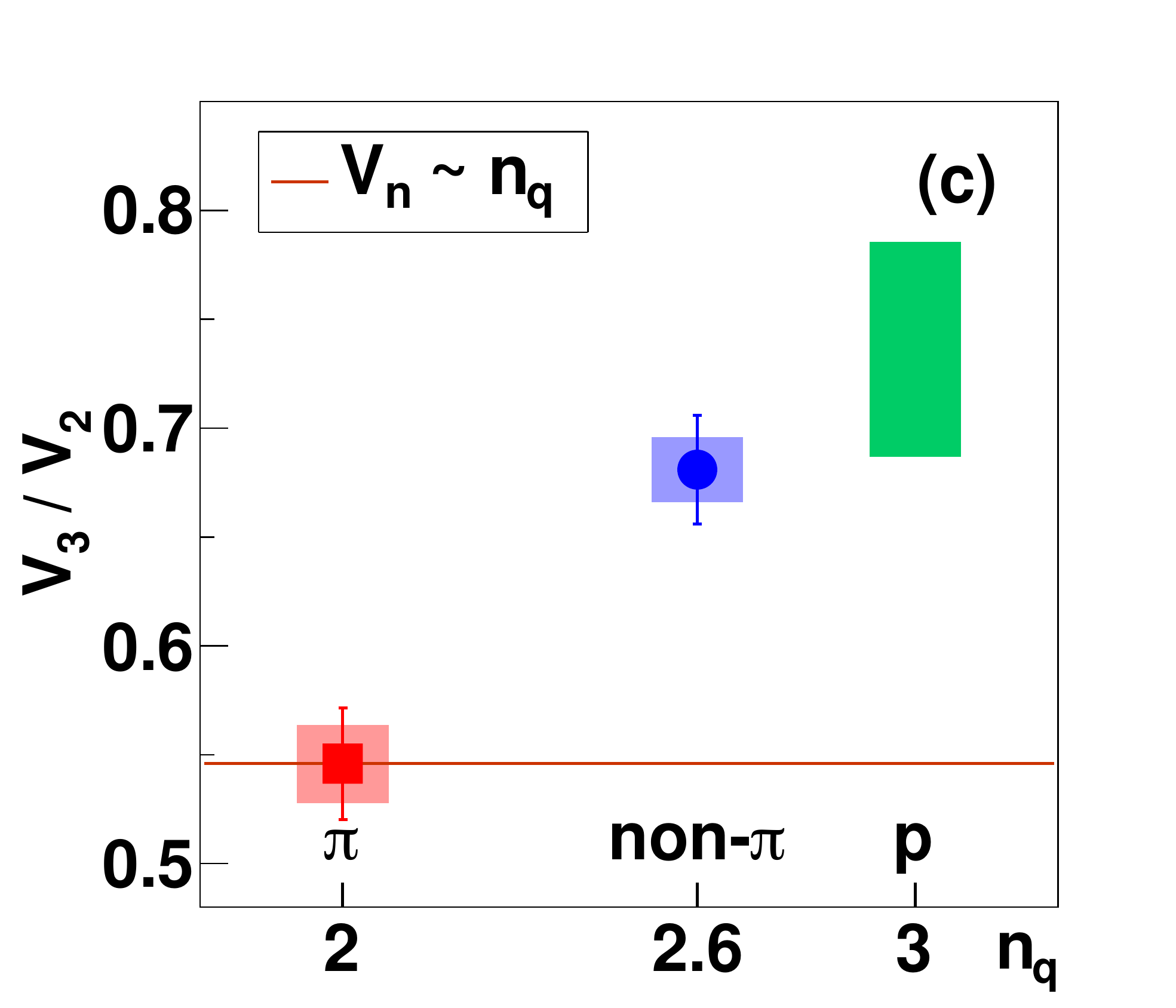}
  \caption{
  (Color online) (a) $\Delta\phi$ projections over $|\Delta\eta|<1.5$ in 0-10\% Au+Au and minimum-bias d+Au data at 200~GeV after subtracting the jet-like fit components.
  Overlapping dashed and dotted lines illustrate the results of the two-dimensional fits for the flow- and mini-jet based models, respectively.
  Only statistical errors are shown.
  (b) Solid symbols show extracted Fourier coefficients from fitting in $\left(\frac\pi2<\Delta\phi<\frac{3\pi}2\right) \times \left(|\Delta\eta|<1.5\right)$ (1.52 for d+Au),
  for pion, non-pion, and charged hadron triggers in the flow-based  model.
  Open symbols show d+Au data, scaled by the ratio of background levels.
  (c) $V_3/V_2$ ratio  for pion and non-pion triggers in Au+Au, and the extrapolated value for ``pure protons'', as described in the text. 
  In all panels, statistical errors are shown as lines (smaller than symbol size for some points), and systematic uncertainties as colored boxes in panels (b) and (c).
}
  \label{fig:Vn}
\end{figure*}

Outside of the jet-like cone region,
we find no $\Delta\eta$-dependence in the correlated yields within our fiducial  range. 
To characterize the long-range contributions in Au+Au,
we perform two-dimensional fits to the full correlation with two different models.
One model attributes the ridge to modified fragmentation of produced mini-jets, 
and the other explains it in terms of higher-order hydrodynamic flows. 
In both models, the near-side jet-like peak is mathematically characterized by a two-dimensional generalized Gaussian
  $e^{-(|\Delta\phi|/\alpha_{\phi})^{\beta_{\phi}}}\,e^{-(|\Delta\eta|/\alpha_{\eta})^{\beta_{\eta}}}$.
The resulting fit parameters for the jet cone are found to be identical between the two models
and were used for extrapolation of the jet-like cone yields presented in Table~\ref{tab:yields}. 

The $\Delta\phi$ projections of the pseudorapidity-independent parts of the two-dimensional correlations (after subtracting the jet-like peak),
are shown in Fig.~\ref{fig:Vn}, panel (a), together with both fit functions discussed below. 
Also shown is the corresponding projection for d+Au data, shifted by an arbitrary offset. We performed the Fourier analysis on this 
data as well; without an appreciable near-side ridge, the above mini-jet model is not applicable here.

In the flow-based approach,
based on a hydrodynamic expansion of an anisotropic medium,
all $\Delta\eta$-independent parts of the correlations are described via Fourier expansion:
$A (1 + \sum_{n=1}^{N} 2 V_n\cos n\Delta\phi )$, where $A$ describes the magnitude of the uncorrelated background,
$V_2$ is conventionally associated with ``elliptic flow'', and $V_3$ with ``triangular flow''.
In this work, the first five terms ($N$=5) exhaust all features of the correlation
to the level of statistical uncertainty, and $V_n$  represents the combined trigger and associated hadron anisotropy parameters.
We note that in this approach, the fragmentation contributions to the
away-side correlations in central Au+Au data are expected to be
strongly suppressed relative to flow effects by quenching, and they
are therefore
neglected~\cite{Alver:2010gr,*Alver:Erratum,Luzum:2010sp}. In the d+Au
data, on the contrary, the away-side jet contributions dominate and
no appreciable near-side correlated yield at large $\Delta\eta$ is present.

The Fourier fit results are shown in Fig.~\ref{fig:Vn} (b).
In Au+Au, the second harmonic is dominant in all long-range correlations
for the central data, followed by the triangular ($V_3$) term.
Higher-order harmonic amplitudes rapidly decrease.
All harmonic amplitudes for non-pion triggers are found to be larger than those for pion triggers, which is qualitatively consistent with recombination expectations.

The corresponding Fourier harmonics in d+Au are found to be consistent with
expectations for a decomposition of a Gaussian peak at $\pi$,
i.e. rapidly falling and with alternating signs. For $n=3$, the harmonics are already consistent with zero within errors, 
confirming that the Fourier and mini-jet approaches are indistinguishable in this case.
 
In central Au+Au collisions at RHIC, elliptic flow parameters of identified hadrons have been shown to scale with the number of constituent quarks
$n_\text{q}$, suggesting collective behavior at the partonic level~\cite{STAR:NCQV2,*PHENIX:NCQV2}.
The estimated  baryon/meson ratio for $V_2$ in this analysis is also consistent with 3/2, see below.
We note that in our trigger $p_T$ range, azimuthal anisotropy is approximately independent of $p_T$~\cite{Adams:2004bi},
eliminating the need to address quark momentum dependence.
To test whether this scaling extends to the triangular term, 
we examine the $V_3/V_2$ ratios.
This test assumes that the measured Fourier coefficients factorize into 
$V_n=\langle v_n^\text{trig}\rangle \langle v_n^\text{assc}\rangle $, where  $v_n^\text{trig}$ and $v_n^\text{assc}$
measure azimuthal anisotropies of trigger and associated hadrons, respectively~\cite{Luzum:2010sp}. 
The factorization has been demonstrated experimentally for $V_2$ in~\cite{Aamodt:2011by}.
The extracted $V_2$ coefficients are found consistent with the product of previously measured
identified and unidentified $v_2$ values. 
Since  the selection of associated particles is identical for all correlations in this analysis,
the anisotropy contributions from associated hadrons should cancel in the ratios of $V_n$ coefficients.
Figure~\ref{fig:Vn} (c) shows $V_3$/$V_2$ ratios extracted from long-range correlations versus average $n_{\text{q}}$ per particle 
for pion and non-pion triggers. The systematic uncertainty, determined by varying the fitting range
and the $\dif E / \dif x$ cut position for pion/non-pion separation, 
was found to be similar to, or smaller than, the statistical uncertainty.
We find that the ratio of triangular and elliptic flow is $0.546\pm0.025$(stat.)$\pm0.018$(sys.)
for pion triggers and $0.681\pm0.025$(stat.)$\pm0.015$(sys.) for non-pions.
If the measured final-state azimuthal anisotropies are indeed of collective partonic origin which transform into final-state hadronic
observables through the coalescence/recombination of constituent quarks, then we would expect the same dependence of all $v_n^\text{trig}$ on constituent quark number. 
Even with the significant meson contribution to non-pion triggers, the ratios give a strong indication
of a breaking of the simple $n_{\text{q}}$ scaling behavior between the second and third Fourier harmonics.
Assuming that kaons, as mesons, adhere to the pion scaling trend, and using the known p/$\pi$ ratio reported in refs.~\cite{Adams:2006wk,*Adams:2006nd},
we construct an estimate of the  $V_3$/$V_2$  ratio 
for pure protons in Fig.~\ref{fig:Vn} (c). The systematic uncertainty in the estimated ``pure-proton'' $V_3/V_2$
value of $0.736\pm0.038$(stat.)$\pm0.032$(sys.)  includes an
additional 1\%  uncertainty from PID.
Feed-down protons from $\Lambda$ closely preserve the original parent direction,
and we expect no measurable effect on the $\Delta\eta$-independent terms, 
as $\Lambda$ and protons have very similar azimuthal anisotropy in our kinematic range.

The observed violation of constituent quark number scaling for $V_3$,
based on the ``pure proton'' extrapolated value for $V_3$, 
$V_3(\text{baryon})/V_3(\text{meson}) = 2.03 \pm 0.12 \text{(stat.)} \pm 0.20 \text{(sys.)}$
compared to 
$V_2(\text{baryon})/V_2(\text{meson}) = 1.50 \pm 0.06 \text{(stat.)} \pm 0.07 \text{(sys.)}$,
is intriguing because recombination/coalescence models are the only ones
presently capable of describing constituent quark scaling behavior among the
$V_2$ parameters for many identified hadrons.


The difference between the $V_3$ and $V_2$ scaling behavior demonstrated in Fig.~\ref{fig:Vn} (c) therefore suggests the
need for other contributions to long-range correlations to explain the data.
We note that deviations from $n_\text{q}$ scaling of elliptic flow have been observed at the LHC~\cite{Abelev:2014pua},
and at RHIC for non-central collisions~\cite{Adare:2012vq}.
The $v_n$ scaling proposed in ref.~\cite{Lacey:2011av} better describes our
measured $V_3/V_2$ ratios, but still under-predicts the enhancement for non-pion triggers.

\begin{table} [!htb]
  \centering
  \caption{First and second harmonic extracted using the mini-jet model in 0-10\% most central Au+Au data at 200~GeV. 
    Note that the amplitudes are multiplied by 100.
    \label{tab:estruct}}
  \begin{ruledtabular}
    \begin{tabular}{  c || c | c }
      {\bf Trigger } &  $100(\,V_1 \pm \text{stat.} \pm \text{sys.}) $  &  $100(\,V_2 \pm \text{stat.} \pm \text{sys.})$    \\ \hline
      $\pi$              & $-0.86 \pm 0.09 \pm 0.07$                            &  $-0.017 \pm 0.045 \pm 0.034$                         \\ \hline 
      non-$\pi$      & $-1.53 \pm 0.13 \pm 0.08$                            &  $-0.343 \pm 0.059 \pm 0.041$                         \\ \hline 
      all                   & $-1.19 \pm 0.05 \pm 0.01$                            &  $-0.173 \pm 0.025 \pm 0.004$                      
    \end{tabular}
  \end{ruledtabular}
\end{table}
 
In the mini-jet model, in which the major component is in-medium modification of fragmentation,
only the first two terms ($N$=2) of the Fourier expansion are kept and
the near-side ridge in this analysis is modeled by a one-dimensional Gaussian,
resulting in $A (1 + 2 V_1\cos \Delta\phi + 2 V_2\cos 2\Delta\phi) + B \,e^{-\Delta\phi^2/2\sigma^2}$.
Here $A$ is the uncorrelated yield, $B$ is the ridge amplitude, and $\sigma$ is the ridge width parameter.
The dipole $V_1$ is designated to describe the away-side jet and/or momentum conservation effects,
and $V_2$ describes a non-jet quadrupole (potentially of flow origin). 
The addition of the 1D near-side Gaussian, which differs from the original model elements in ref.~\cite{STAR:Anomalous},
was necessary to reproduce the data, as noted in ref.~\cite{Ray2013}.
The mini-jet model fit describes the measured Au+Au correlations for all three trigger types as well as the
flow-based approach (Fig.~\ref{fig:Vn} (a)),
yielding identical uniformly distributed residuals and $\chi^2$ values.
The extracted harmonic amplitudes are shown in Table.~\ref{tab:estruct}.
As the away-side structure is for the most part described by the dipole term,
the magnitude of the $V_1$ amplitude is found to be significantly larger for leading non-pions than for pions.
For back-to-back jets, this $V_1$ increase is supposed to balance the near-side (leading) jet contributions,
which would have to consist of both the jet-like peak and the ridge
because the jet-like peak alone decreases for non-pion
trigger particles.
Understanding the behavior of the $V_2$ term in the mini-jet model fits is challenging:
the $V_2$ amplitude, while consistent with zero for pion triggers, is significantly negative for non-pion triggers. 
This negative value for $V_2$, which is conventionally associated with elliptic
flow, is not expected from any known source and calls into question the
applicability of the assumed parameterization for the centrality
and $p_T$ range studied here, the validity of the ``mini-jets + quadrupole only'' physics scenario,
or both.

In summary, a statistical separation of pion and non-pion triggers was performed to study the systematic behavior  of  di-hadron correlations 
from central Au+Au and minimum-bias d+Au collisions at 200 GeV with the STAR experiment.   
The correlations, decomposed into short- and long-range parts in $\Delta\eta$, are analyzed for different identified trigger types
to test the consistency of two models in order
to improve our understanding of hadronization mechanisms in the quark gluon plasma. 
We find significant enhancement of intermediate-$p_T$ charged-hadron jet-like yields 
associated with pion triggers relative to a d+Au reference measurement.
%
The enhancement is qualitatively consistent with observed modifications of jet fragmentation functions measured at the LHC,
suggesting it results from the energy loss process.
For the non-pion trigger sample, a larger contribution from gluon fragmentation is expected 
compared to pion triggers~\cite{Albino:AKK,Adams:2006nd,Mohanty2007}.
Due to the color-charge factor, a larger energy loss for gluons is expected relative to that of quarks.
No enhancement is observed for non-pion triggers in contrast to pQCD-based
expectations for color charge dependence of energy loss.
This lack of enhancement may indicate a competition between parton-medium interaction effects
and dilution of jet triggers by quark recombination contributions. 

No statistically significant ridge is found associated with either trigger type in minimum bias d+Au data.
In Au+Au data, we find a significantly larger ridge-like yield and away-side correlation strength for non-pion than for pion triggers. 
Two fitting models which are mathematically similar but which are based on distinct physical assumptions were applied to the Au+Au data.
Both models, while describing the correlations well,
attain parameter values which are problematic within the assumed physical scenarios. 
In the flow model, the observed differences of  $V_3/V_2$ ratios
imply that the explanation of the ridge and away-side modifications as resulting 
only from hydrodynamic flow of a partonic medium with constituent quark recombination at hadronization is incomplete.
On the other hand, the negative $V_2$ result for the mini-jet based model for leading non-pions indicates that for the
data reported here, either the assumed scenario or the mathematical parameterization for jets and dijets is inadequate, or both.
These results may have significant implications for understanding the origin of the ridge and hadronization in the QGP.

We thank the RHIC Operations Group and RCF at BNL, the NERSC Center at LBNL, the KISTI Center in Korea, and the Open Science Grid consortium for providing resources and support. This work was supported in part by the Offices of NP and HEP within the U.S. DOE Office of Science, the U.S. NSF, CNRS/IN2P3, FAPESP CNPq of Brazil,  the Ministry of Education and Science of the Russian Federation, NNSFC, CAS, MoST and MoE of China, the Korean Research Foundation, GA and MSMT of the Czech Republic, FIAS of Germany, DAE, DST, and CSIR of India, the National Science Centre of Poland, National Research Foundation (NRF-2012004024), the Ministry of Science, Education and Sports of the Republic of Croatia, and RosAtom of Russia.

\bibliography{../tex/RHIC}

\end{document}